\documentclass[12pt,a4paper]{article}
\usepackage{amssymb}
\usepackage{epsf}
\usepackage{cite}
\newlength{\HFPP}       \HFPP5.4mm
\makeatletter
\@addtoreset{equation}{section}
\makeatother

\def\preprint#1#2{\noindent\hbox{#1}\hfill\hbox{#2}\vskip 10pt}
\def\k_tilde{\sqrt{\frac{\kappa}{1-\kappa}}}
\def\kt2{\frac{\tilde{\kappa}}{2}}

\begin{document}
\begin{titlepage}
\def\thefootnote{\fnsymbol{footnote}}

\preprint{ITP-UH-25/98}{November 1998}
\vfill

\begin{center}
  {\Large\sc A generalized spin ladder in a magnetic field}
\vfill

{\sc Holger Frahm}\footnote{e-mail: frahm@itp.uni-hannover.de}
	and
{\sc Claus R\"odenbeck}
\vspace{1.0em}

{\sl
  Institut f\"ur Theoretische Physik, Universit\"at Hannover\\
  D-30167~Hannover, Germany}
\end{center}
\vfill

\begin{quote}
We study the phase diagram of coupled spin-$1/2$ chains with bilinear
and (chiral) three-spin exchange interactions in a magnetic field.
The model is soluble on a one-parametric line in the space of coupling
constants connecting the limiting cases of a single and two decoupled
Heisenberg chains with nearest neighbour exchange only.  We give a
complete classification of the low-energy properties of the integrable
system and introduce a numerical method which allows to study the
possible phases of spin ladder systems away from the soluble line in a
magnetic field.
\end{quote}


{PACS-Nos.: 
75.10.Jm	
75.30.Kz	
75.50.Ee	
}
\vspace*{\fill}
\setcounter{footnote}{0}
\end{titlepage}

%
\section{Introduction}
One dimensional quantum spin systems have attracted much interest in
recent years.  On the theoretical side, soluble models and powerful
methods such as bosonization combined with numerical results have been
used to understand many aspects of the rich physics found in these
systems.  In particular the Bethe Ansatz solution of the $S=1/2$
Heisenberg chain \cite{bethe:31} with nearest neighbour interaction
has provided much insight into their properties.
At the same time the fabrication of materials which are essentially
quasi-one dimensional spin-$1/2$ systems allows for experimental
studies of these features.
In addition to nearest neighbour exchange the effects of competing
next-nearest neighbour interactions driving a spin Peierls transition
\cite{cace:95,riko:95,whaf:96} and interchain interaction as in spin
ladders \cite{bdrs:93,gors:94,shex:96,brmn:96} have been considered to
account for the properties of these compounds.  Furthermore multi-spin
exchange terms can be included without breaking the $SU(2)$ symmetry
of the system \cite{nets:97,komi:98,komi:98b}.  In general such terms
are known to drive phase transitions in the Heisenberg chain opening a
gap $\Delta$ between the singlet ground state and the lowest (triplet)
excitation.

New interesting properties can be observed when such systems are
placed in a magnetic field \cite{chgi:97,ussu:98}:
For $H>\Delta/g\mu_B$ the system undergoes a transition from the
gapped phase to an incommensurate phase with non-zero magnetization in
which it has gapless excitations until the field is strong enough to
lead to a saturated ferromagnetic state.  Signatures of these
transitions and properties of the gapless high-field phase have
already been studied in several organic $S=1/2$ spin Peierls and
ladder systems \cite{Kir95,Chab97,Chab98}.

The subject of this paper is to study of the properties of this
gapless phase.  Our starting point is a soluble model of coupled
spin-$1/2$ Heisenberg chains ($0\le\kappa\le1$)
\cite{pozv:93,zvya:94,frcr:96,frcr:97}
\begin{eqnarray}
{\cal H} &=& \sum_{i=1}^N \left\{J_1
    \left(\vec{S}_{2i-1}\vec{S}_{2i} 
	+ \vec{S}_{2i} \vec{S}_{2i+1}\right) 
 +J_2\,\left(\vec{S}_{2i}\vec{S}_{2i+2}
               +\vec{S}_{2i+1}\vec{S}_{2i+3}\right) 
                                          \right.
\nonumber \\
         & &    \left.   +J_3
 \left(\vec{S}_{2i-1}\left(\vec{S}_{2i} \times \vec{S}_{2i+1}\right)
     - \vec{S}_{2i}\left(\vec{S}_{2i+1} \times \vec{S}_{2i+2}\right)
 \right)\right\}
\label{hamil}\\[4pt]
  && J_1 = 2J(1-\kappa)\ ,\quad
  J_2 = J\kappa\ ,\quad
  J_3 = 2J\sqrt{\kappa(1-\kappa)}\ .
\nonumber
\end{eqnarray}
The three-spin exchange term guarantees integrability of the system
for arbitrary ratio of antiferromagnetic pair exchange constants,
i.e.\ $0<\kappa<1$ (for more general cases see e.g.\ \cite{pale:98}).
It breaks parity and time-reversal invariance, for a possible
mechanism for the appearence of such a term see Ref.~\cite{Okamoto96}.
Below we put $J=1$ which fixes our scale of energy.  In the following
section we use the Bethe Ansatz solution to give a detailed account of
the magnetic phase diagram and the critical properties of this model.
Unlike the systems discussed above the model (\ref{hamil}) has gapless
excitations for \emph{any} $\kappa$ as long as the magnetization is
not saturated.  Furthermore, for sufficiently large $\kappa$ we find a
transition into a \emph{second} critical phase with different
universality class at an intermediate magnetic field \cite{frcr:97}.
Based on these findings we introduce a numerical method for the
determination of the phase boundaries which allows to extend this
discussion to \emph{non}integrable generalizations of (\ref{hamil}).
Finally we apply this method to propose a phase diagram of the system
with competing nearest and next-nearest neighbour exchange (i.e.\
without the three spin terms in (\ref{hamil})).

%
\section{The integrable model}
\label{sec:int}
Starting from the ferromagnetic eigenstate with all spins up the
spectrum of the integrable Hamiltonian (\ref{hamil}) is obtained by
adding magnons which are parametrized by the roots of the
Bethe--Ansatz equations (BAE) \cite{pozv:93}
\begin{equation}
\left(\frac{\lambda_j - \tilde{\kappa} + \frac{i}{2}}
           {\lambda_j - \tilde{\kappa} - \frac{i}{2}}\right)^N
\left(\frac{\lambda_j + \tilde{\kappa} + \frac{i}{2}}
           {\lambda_j + \tilde{\kappa} - \frac{i}{2}}\right)^N
    = \prod_{j\not= k}^M
           \frac{\lambda_j-\lambda_k+i}{\lambda_j-\lambda_k-i}\ 
\label{BAE}
\end{equation}
with $2\tilde{\kappa}=\sqrt{\kappa/(1-\kappa)}$.  The corresponding
state has magnetization ${\cal M} = N-M$ and energy
\begin{equation}
  E(\{\lambda_j\}) = \pi \sum_{j=1}^M \left(
	a_1(\lambda_j +\tilde{\kappa})
       +a_1(\lambda_j -\tilde{\kappa})\right)\ ,
\quad
  \pi a_n(\lambda) = -\frac{2n}{4\lambda^2+n^2}\ . 
\label{Elam}
\end{equation}
Generic solutions to the BAE are arranged in $m$-strings of complex
rapidities $\lambda_j^{(m)} = x+i((m+1)/2-j)$, $j=1,\ldots,m$ with
common real part.  For the ground state only \emph{real} solutions to
(\ref{BAE}) have to be taken into account.  
In the thermodynamic limit the ground state in a magnetic field
$H=h/g\mu_B$ is obtained by filling all states with negative
\emph{dressed} energy, defined in terms of the integral equation
\begin{equation}
  \epsilon_1(\lambda)
	+ \int_{\epsilon(\lambda)<0}d\mu\
		a_2(\lambda-\mu) \epsilon(\mu)
  = h + \pi\left(a_1(\lambda+\tilde\kappa)
		  +a_1(\lambda-\tilde\kappa) \right)\ .
\label{dressE}
\end{equation}
For $h=0$ one finds $\epsilon_1<0$ for all $\lambda$, the
corresponding ground state is parameterized uniquely by $N$ real
rapidities, giving a singlet state with energy ($\Psi(x)$ is the
digamma function)
\begin{equation}
  \frac{E}{N}    
    = -\ln 2 
      -\frac{1}{4}\left\{\Psi\left(1-i\tilde{\kappa}\right) 
      -\Psi\left(\frac{1}{2}-i\tilde{\kappa}\right) 
      +\Psi\left(1+i\tilde{\kappa}\right) 
      -\Psi\left(\frac{1}{2}+i\tilde{\kappa}\right)\right\}\ .
\label{gse}
\end{equation}
The low-lying excitations are scattering states of spinons
\cite{fata:84} and the low energy behaviour of the theory is that of a
level-1 $SU(2)$ Wess-Zumino-Witten model with central charge $c=1$
independent of $\kappa<1$ \cite{affl:90}.
A small magnetic field breaks the $SU(2)$ symmetry giving a $c=1$
Gaussian conformal field theory (CFT) with anomalous dimensions
depending on $h$.  Increasing the magnetic field $h$ beyond
\begin{equation}
  h \ge h_{c2} = \left\{
	\begin{array}{ll}
	4(1-\kappa)&\mathrm{for~}0\le\kappa\le{1\over4}\\
	1+\kappa^{-{1\over2}}&\mathrm{for~}{1\over4}\le\kappa\le1
	\end{array}
\right. 
\label{hc2}
\end{equation}
one has $\epsilon_1(\lambda)>0$ resulting in a ferromagnetically
polarized ground state \cite{frcr:97}.

For $\kappa >1/4$ an intermediate phase can be found for sufficiently
strong fields $h_{c1}<h<h_{c2}$ which is characterized
by
\begin{equation}
  \epsilon_1(\lambda) <0 \hbox{~for~} \Lambda_1<|\lambda|<\Lambda_2\
\end{equation}
($\Lambda_{1,2}$ have to be determined as functions of the magnetic
field from (\ref{dressE}).)
Here the ground state is formed by filling two 'Fermi seas' of magnons
with positive (negative) rapidities $\lambda$.  The presence of
gapless excitations near the Fermi points at $\lambda=\pm \Lambda_1,
\pm \Lambda_2$ changes the low energy spectrum of the system which in
this region has to be described in terms of \emph{two} $c=1$ Gaussian
CFTs \cite{frcr:97} very similar to a Luttinger liquid of electrons in
one spatial dimension \cite{frko:90,kaya:91}.
The asymptotic behaviour of correlation functions can be determined
from the finite-size corrections to the energy of low lying states.
Following Refs.~\cite{vewo:85,woyn:89} we obtain (see Appendix
\ref{app:fss})
\begin{equation}
\Delta E = -\frac{\pi}{6N^2}(v_1+v_2) + \frac{2\pi}{N^2}\left\{
  v_1(\Delta_1^++\Delta_1^-)+v_2(\Delta_2^++\Delta_2^-)\right\}
\label{fsE}
\end{equation}
where $v_{1,2}$ are the Fermi velocities (\ref{veloF}) of excitations
near the points $\Lambda_{1,2}$ and
\begin{eqnarray}
\Delta_1^\pm &=& \frac{1}{2}\left\{
\frac{(x_{12}\Delta M_2-x_{22}\Delta M_1)}{2({\rm det}\,{\bf X})}
      \mp \frac{(z_{12}\Delta D_2-z_{22}\Delta D_1)}{2({\rm det}\,{\bf
      Z})}\right\}^2\ ,
\nonumber\\
\Delta_2^\pm &=& \frac{1}{2}\left\{
\frac{(x_{11}\Delta M_2-x_{21}\Delta M_1)}{2({\rm det}\,{\bf X})}
      \mp \frac{(z_{11}\Delta D_2-z_{21}\Delta D_1)}{2({\rm det}\,{\bf 
      Z})}\right\}^2
\label{Deltas}
\end{eqnarray}
are the conformal dimensions of the primary fields in the two Gaussian
models expressed in terms of $2\times2$-matrices ${\bf X}$ and ${\bf
Z}$ with elements (\ref{XXX}) and (\ref{ZZZ}) respectively.\footnote{%
In other systems allowing for several types of massless excitations
the relation ${\bf X} \propto {\bf Z}^{-1}$ allows to parameterize the
critical exponents in terms of a single \emph{dressed charge matrix}
\cite{izkr:89,fryu:90,frko:90,kaya:91}.  Contrary to our claim in
Ref.~\cite{frcr:97} this is not possible in this model.}
The integers $\Delta M_i$ and $\Delta D_i$ denote the difference of
the quantum numbers (\ref{MD}) between the excitation and the ground
state.

\begin{figure}[ht]
\begin{center}
\leavevmode
\epsfxsize=0.7\textwidth
\epsfbox{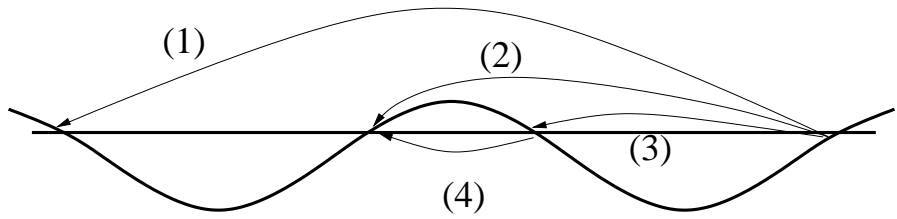}
\end{center}
\caption{Low-energy processes contributing to $C^{zz}(x)$.
\label{fig:excited}}
\end{figure}

\begin{figure}[ht]
\begin{center}
\leavevmode
\epsfxsize=0.6\textwidth
\epsfbox{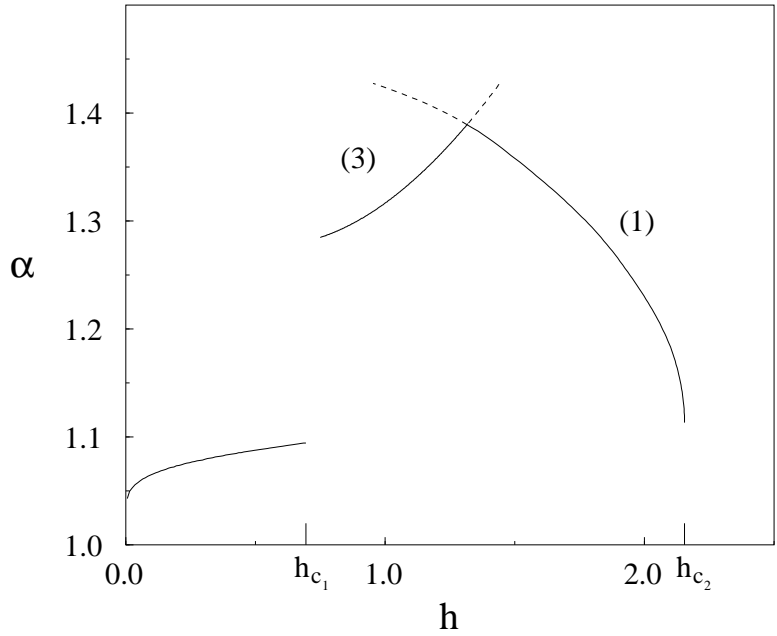}
\end{center}
\caption{Critical exponent $\alpha$ of $C^{zz}(x)$ for $\kappa=3/4$ as
a function of the magnetic field.  The cusp at $h\approx1.31$ is a
consequence of the crossover from the intrachain (3) to the interchain
process (1) in Fig.~\protect{\ref{fig:excited}}.
\label{fig:alpha}}
\end{figure}

Taking the limit $h\searrow h_{c1}$ corresponding to $\Lambda_1=0$ the
matrix elements of ${\bf X}$ and ${\bf Z}$ can be expressed in terms
of the scalar {\em dressed charge} $\xi$ satisfying the integral
equation $\xi(\lambda) = 1 - \int_{-\Lambda_2}^{\Lambda_2} d\mu
a_2(\lambda-\mu)\xi(\mu)$ as
\begin{equation}
{\bf X} = 
	\left( \begin{array}{cc}
	1 & 0 \\
	1-\xi(0) & \xi(\Lambda_2)
	\end{array} \right), \quad
{\bf Z} = 
	\frac{1}{\xi(\Lambda_2)}\left( \begin{array}{cc}
	\xi(\Lambda_2) & 1-\xi(0) \\
	0 & 1
	\end{array} \right).
\end{equation}
This allows to simplify the expressions for the conformal dimensions
(\ref{Deltas}) for $h=h_{c1}+0$ as
\begin{eqnarray}
  \Delta_1^\pm &=& \frac{1}{8}\left\{
	\Delta M_1 \pm \left[(1-\xi(0))\Delta D_2 - \Delta
	D_1\right]\right\}^2
\nonumber \\
  \Delta_2^\pm &=& \frac{1}{8}\left\{
     \frac{(1-\xi(0))\Delta M_1-\Delta M_2}{\xi(\Lambda_2)} 
	\pm \xi(\Lambda_2)\Delta D_2\right\}^2\ .
\label{Deltasc1}
\end{eqnarray}
From (\ref{fsE}) the correlation function of the operator with quantum
numbers $\Delta M_i$, $\Delta D_i$ is found to decay algebraically as
$(1/x)^\alpha$ with an exponent
%
$
\alpha = 2\left(\Delta_1^++\Delta_1^-+\Delta_2^++\Delta_2^-\right)
$ \cite{frko:90,kaya:91}.
%
Due to the nature of the ground state in the regime $h_{c1}<h<h_{c2}$
several low energy processes contributing to $C^{zz}(x)=\langle
S^z(x)S^z(0) \rangle$ have to be considered (see
Fig.~\ref{fig:excited}).  The corresponding quantum numbers in
(\ref{Deltas}) are
\begin{equation}
\begin{array}{lllll}
(1) &\Delta D_1=0 & \Delta D_2 = 2 & \Delta M_1 = 0 &\Delta M_2 =0\\
(2) &\Delta D_1=-1 & \Delta D_2 = 1 & \Delta M_1 =1 &\Delta M_2 =1\\
(3) &\Delta D_1=1& \Delta D_2 = 1 & \Delta M_1 = 1& \Delta M_2 = 1\\
(4) &\Delta D_1=2& \Delta D_2 = 0& \Delta M_1 = 0 &\Delta M_2 = 0
\end{array}
\label{qnos}
\end{equation}
The leading asymptotic behaviour of $C^{zz}(x)$ is determined by the
\emph{smallest} exponent $\alpha$.  Approaching the transition line
$h=h_{c1}(\kappa)$ from above we find that this exponent corresponds
to the process (1) for $1/4<\kappa <\kappa_c (h_{c1})$ and (3) for
$\kappa_c(h_{c1})< \kappa< 1$.  At $\kappa= \kappa_c (h_{c1})
\approx0.61$ or $h_{c1}(\kappa_c)\approx 1.35$ a crossover from an
effective single chain to two chain behaviour takes place.  Comparing
the resulting exponents $\alpha$ to the one obtained in the single
Gaussian theory for $h\nearrow h_{c1}$ \cite{boik:86} we find from
(\ref{Deltasc1})
\begin{equation}
\alpha = \left\{ \begin{array}{ll}
        2\xi(\Lambda_2)^2 &\hspace{2mm} \mbox{} \hspace{5mm}0\leq h<h_{c1} \\
        2(1-\xi(0))^2+2\xi(\Lambda_2)^2  
         &\hspace{2mm} \mbox{}\hspace{5mm} h\searrow h_{c1},\,
        \kappa < \kappa_c(h_{c1})\\
        \frac{1}{2}
        \left(1+\xi(\Lambda_2)^{-2})(\xi(\Lambda_2)^2+\xi(0)^2\right) 
        &\hspace{2mm} \mbox{}\hspace{5mm} h\searrow h_{c1},\, 
        \kappa >\kappa_c(h_{c1})
\end{array}\right.\ 
\end{equation}
This crossover in the long distance asymptotics of $C^{zz}(x)$ in the
high field phase with $h_{c1}<h<h_{c2}$ can be observed for any
$\kappa>\kappa_c(h_{c1})$: in the limit $\kappa\to1$ of decoupled
chains the exponents due to the most relevant processes (1) and (3) in
(\ref{qnos}) are $\alpha^{(1)} = 1/\xi(\Lambda)^2$ and $\alpha^{(3)}=
2\xi(\Lambda)^2$ where $\Lambda=(\Lambda_2-\Lambda_1)/2$.  Again the
dominant process at low energies is determined by the effective
single-chain process with exponent $\alpha^{(3)}$ for small fields
with a crossover to the two-chain process (1) at a magnetic field to
be determined from $2\xi(\Lambda)^4=1$, or $h\approx1.29$.  Solving
the integral equations numerically we obtain the complete phase
diagram of the integrable chain in a magnetic field
(Fig.~\ref{fig:ex}).  Depending on $\kappa$ and $h$ the low energy
properties of the system are those of a single or of two Gaussian
CFTs.  Increasing the magnetic field in the latter regime a crossover
is observed corresponding a change from the \emph{intra}chain process
(3) to the \emph{inter}chain process (1) as the most important one at
low energies.  The typical dependence of the exponent $\alpha$ for
fixed ratio of the exchange constants on the magnetic field is shown
in Fig.~\ref{fig:alpha}.

%
%
\section{Phase diagram of generalized spin ladders}
\label{sec:num}
\begin{figure}[ht]
\begin{center}
\leavevmode
\epsfxsize=0.6\textwidth
\epsfbox{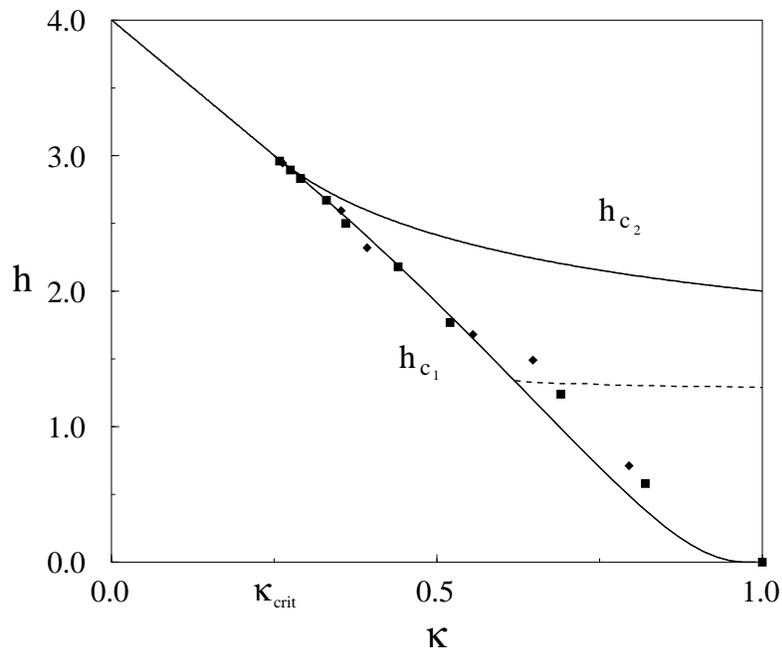}
\end{center}
\caption{Magnetic phase diagram of the integrable model.  The dashed
line marks the crossover between the intrachain and interchain process
as dominant low energy contribution to $C^{zz}$, the symbols denote
the position of the phase transition as determined from numerical
diagonalization of systems with $16$ and $20$ spins using the
singularities in the operlaps (\protect{\ref{num:ov}}).
\label{fig:ex}}
\end{figure}

To investigate whether the magnetic phases found in the integrable
system (\ref{hamil}) persist when we relax the conditions on the pair
and three spin exchange constants we consider the Hamiltonian
(\ref{hamil}) with general values of the exchange constants $J_i$.
For magnetic fields strong enough to polarize the system completely
(corresponding to $h_{c2}$ in the previous section) the excitations
above the ferromagnetic ground state are spin waves.  Their dispersion
is easily obtained to be
\begin{equation}
  E_\pm = J_2\cos k \pm
	\sqrt{J_1^2\cos^2(k/2) + (J_3^2/4)\sin^2 k} + \mathrm{const.}
\label{num:Epm}
\end{equation}
Below $h_{c2}$ the ferromagnetic state becomes unstable against
creation of these magnons.  Depending on the coupling constants $J_i$
one finds either a single magnon mode at $k=0$ or two modes with
wave numbers $\pm Q$ becoming soft at the transition into the
paramagnetic phase---just as in the integrable model discussed in the
previous section.  The transition between these two scenarios occurs
at the point where the low energy magnon modes (\ref{num:Epm}) have a
quartic dispersion $E_-\sim k^4$ for small wave numbers.  This requires
a ratio 
\begin{equation}
   {J_2\over J_1} = {1\over4}\left(1-{J_3^2\over J_1^2}\right)
\label{qudisp}
\end{equation}
for the pair exchange constants.  For the integrable case
(\ref{hamil}) this is the point $(\kappa,h)=(1/4,3)$ in the phase
diagram Fig.~\ref{fig:ex}, i.e.\ the upper end point of the transition
line between the two gapless phases.  For a model \emph{without} three
spin exchange terms $(J_1,J_2,J_3)=(2(1-\kappa),\kappa,0)$
Eq.~(\ref{qudisp}) implies a similar transition for $\kappa=1/3$ or
$J_2/J_1=1/4$ (see also below).

For magnetic fields $h<h_{c2}$ computation of spin wave dispersions as
in (\ref{num:Epm}) is not sufficient due to strong quantum
fluctuations in one spatial dimension.  A possible criterion to
determine the phase boundary at smaller fields is the field dependence
of the zero temperature magnetization which shows a characteristic
singularity on the transition line \cite{frcr:97}.  However, while
this feature can be identified easily in the thermodynamic limit
accessible for the Bethe Ansatz soluble model (\ref{hamil}) it cannot
be used to analyze the numerical data due to the discrete set of
magnetizations realized in a finite size system.  Similarly, the
dramatic changes in the low energy spectrum from a single Gaussian
model to the form (\ref{fsE}) are difficult to see from the finite
size data obtained by numerical diagonalization.

\begin{figure}[ht]
\begin{center}
\leavevmode
\epsfxsize=0.6\textwidth
\epsfbox{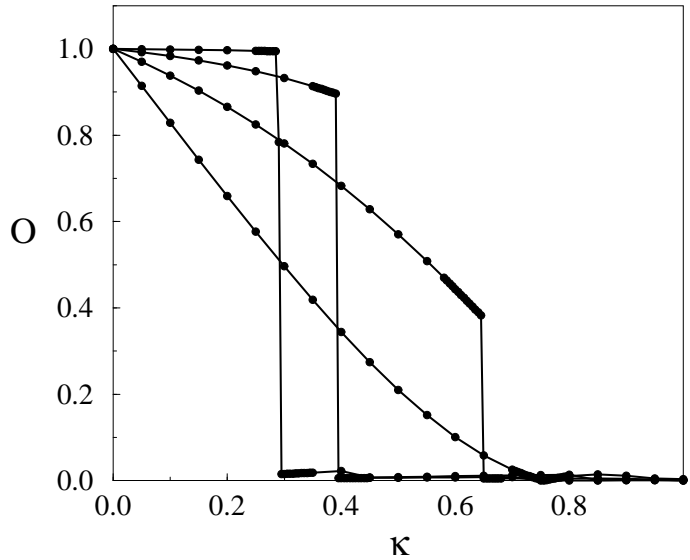}
\end{center}
\caption{Overlap matrix elements (\protect{\ref{num:ov}}) for the
integrable model with 16 spins as a function of $\kappa$:  the data at
small $\kappa$ correspond to states with magnetization $M^z=6,4,2,0$
(top to bottom), the lines are guides to the eye.
\label{fig:cross16}}
\end{figure}

As a possible method to identify the transition line $h_{c1}$ from
numerical finite size data we propose the following method:  having
identified the phase for \emph{small} ratios $J_2/J_1$ as similar in
nature with the single Heisenberg chain $J_2=0=J_3$ we compute the
overlap
\begin{equation}
  {\cal O}(J_1,J_2,J_3) \equiv \left|
	\langle \left\{J_1,0,0\right\}|
  		\left\{J_1,J_2,J_3\right\}\rangle\right|
\label{num:ov}
\end{equation}
between the finite size ground states $|\left\{J_1,J_2,J_3\right\}
\rangle$ of (\ref{hamil}) with fixed magnetization and the
corresponding one of the Heisenberg chain.  Clearly, this quantity
will decrease from the value ${\cal O}(J_1,0,0)=1$ for increasing next
nearest neighbour and three spin interactions.  In
Fig.~\ref{fig:cross16} we present data for (\ref{num:ov}) obtained for
the integrable system (\ref{hamil}) with $16$ spins in the sectors
with different magnetization showing a sharp transition as a function
of $\kappa$.  As shown in Fig.~\ref{fig:ex} these transitions provide
an excellent numerical estimate of the critical field $h_{c1}$.

\begin{figure}[ht]
\begin{center}
\leavevmode
\epsfxsize=0.6\textwidth
\epsfbox{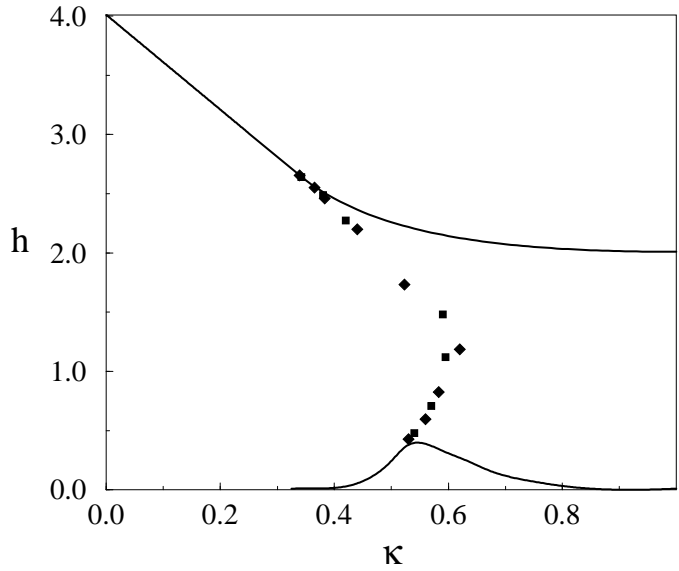}
\end{center}
\caption{Phase diagram of the spin chain model with nearest and
next-nearest neighbour pair exchange $2J_2/J_1=\kappa/(1-\kappa)$: the
upper line denotes the saturation field $h_{c2}$ obtained from spin
wave calculation, the lower one denotes the transition at $h=\Delta$
into the dimerized phase with a spin gap ($0.325\lesssim\kappa<1$ for
$h\to0$).  Squares (diamonds) denote the transition determined
numerically from the overlap criterion (\protect{\ref{num:ov}}) for
systems with $20$ ($16$) spins.
\label{fig:j1j2}}
\end{figure}

Away from the integrable line this sharp transition of the overlap
(\ref{num:ov}) as a function of $J_2/J_1$ is smoothed out.  However,
there are still well defined inflection points of the function ${\cal
O}(J_1,J_2,J_3)$ along lines in the space of parameters $J_{2,3}$.
This allows to estimate the position $h_{c1}$ of the transition in the
\emph{non}integrable model.  In Fig.~\ref{fig:j1j2} the resulting
phase diagram is shown for the system with nearest and next nearest
neighbour pair exchange only, i.e.\ $(J_1,J_2,J_3)
=(2(1-\kappa),\kappa,0)$.  Without an external magnetic field this
model is known to have gapless excitations for $0\le J_2/J_1\lesssim
0.241$ and in the limit of decoupled chains $J_1\to 0$
\cite{cace:95,riko:95,whaf:96}.  For intermediate values of the next
nearest neighbour exchange the system has a spin gap $\Delta$ leading
to a plateau in at $M^z=0$ in the magnetization curve extending to
$h=\Delta$.  For strong magnetic fields $h\sim h_{c2}$ the spin wave
result (\ref{qudisp}) shows a transition from a phase with one and two
types of magnons with quadratic dispersion, respectively.  For
intermediate fields the $\kappa$-dependence of ${\cal
O}(2(1-\kappa),\kappa,0)$ allows to locate the transition between
these phases which is found to end near $\kappa\approx0.5$ at the
transition into the gapped phase.

%
\section{Summary and Conclusion}
To summarize we have presented a detailed account of the magnetic
phase diagram of an integrable model (\ref{hamil}) of generalized
coupled spin chains.  We have established two distinct zero
temperature phases which can be described in terms of a single and two
Gaussian conformal field theories respectively.  The exact finite-size
corrections in the spectra of low energy excitations in the latter
phase allow to classify the possible critical exponents arising in the
long distance asymptotics of two-point correlation functions.  As an
example the exponent in $C^{zz}(x)$ has been found to show a strong
dependence on the magnetic field.  In the same correlator we have
found a crossover from an intrachain to an interchain process as the
most relevant one at low energies.

Finally we have extended our discussion to the possible magnetic
phases of the system (\ref{hamil}) with general exchange constants, in
particular the spin chain with competing nearest and next-nearest pair
exchange only.  Combining known results on the nature of the zero
field ground state with spin wave calculations and numerical
simulations using the overlap (\ref{num:ov}) as a criterion for the
transition we have obtained the phase diagram of this system.
Assuming that the phase found at large $\kappa$ is of similar nature
as in the integrable model (this is certainly true in the large field
limit) this gives strong constraints on a proper bosonization of the
model in the gapless phase found for $h>\Delta$.  We should note,
however, that by construction our analysis can not exclude the
existence of additional phases at larger values of $J_2/J_1$ (which
are not present in the integrable model).  An study in the limit of
two weakly coupled chains ($J_1\ll J_2$) in a magnetic field may give
additional insights on this region.

\section*{Acknowledgments}
This work has been supported by the Deutsche Forschungsgemeinschaft
under Grant No.\ Fr~737/2--3.

\appendix
%
\section{Finite size spectrum of the integrable model}
\label{app:fss}
To compute the finite-size spectrum of the integrable model in the
regime $h_{c1}<h<h_{c2}$ we consider solutions of the BAE (\ref{BAE})
where all possible \emph{real} solutions in the intervals ${\cal I}_-=
[\Lambda_2^-,\Lambda_1^-]$ and ${\cal I}_+= [\Lambda_1^+,\Lambda_2^+]$
with $\Lambda_1^-<\Lambda_1^+$ are present (see
Refs.\cite{vewo:85,woyn:89}).  Their density satisfies the equation
\begin{equation}
  \rho(\lambda)
  + \left\{\int_+ + \int_-\right\}d\mu\ a_2(\lambda-\mu) \rho(\mu)
  = a_1(\lambda+\tilde\kappa) + a_1(\lambda- \tilde\kappa)\ .
\end{equation}
where $\int_\pm$ denotes integration over the interval ${\cal I}_\pm$.
From $\rho$ we define the quantum numbers (related to total
magnetization and momentum) of the corresponding state as
\begin{equation}
  m_i = {M_i\over N} =
        \int_{\Lambda_i^-}^{\Lambda_i^+}d\mu\,\rho(\mu)\ ,
\qquad
  d_i = {D_i\over N} = \left(
	\int_{-\infty}^{\Lambda_i^-} -\int_{\Lambda_i^+}^\infty
	\right)d\mu\, \rho(\mu)\ .
\label{MD}
\end{equation}
To express the finite size corrections to the energy of this state in
terms of these quantum numbers, the matrix $\partial\{m_i,d_i\}/
\partial\{\Lambda_j^+,\Lambda_j^-\}$ has to be computed.  Defining
function $g(\lambda|\Lambda)$ with
\begin{equation}
  g(\lambda|\Lambda) + 
  \left\{\int_+ + \int_-\right\}d\mu\ a_2(\lambda-\mu)g(\mu|\Lambda)
  = a_2(\lambda-\Lambda)\
\label{intG}
\end{equation}
and restricting ourselves to the case $\Lambda_i^\pm=\pm\Lambda_i$
relevant for the Hamiltonian (\ref{hamil}) we obtain
\begin{equation}
 x_{ij} \equiv \pm {1\over\rho(\Lambda_j)}\,
   \frac{\partial m_i}{\partial \Lambda_j^\pm}
   = \delta_{ij} - (-1)^j\int_{-\Lambda_i}^{\Lambda_i} d\mu\,
	g(\mu|\Lambda_j)\ 
\label{XXX}
\end{equation}
and
\begin{equation}
 z_{ij} \equiv {1\over\rho(\Lambda_j)}\,
   \frac{\partial d_i}{\partial \Lambda_j^\pm} =
   \delta_{ij} - (-1)^j \left(
	\int_{-\infty}^{-\Lambda_i} -\int_{\Lambda_i}^\infty
	\right)d\mu\, g(\mu|\Lambda_j)\ .
\label{ZZZ}
\end{equation}
Following Woynarovich \cite{woyn:89} these identities yield (\ref{fsE}) and
(\ref{Deltas}) with the Fermi velocities
\begin{equation}
  v_i = (-1)^i {1\over2\pi \rho(\Lambda_i)}\,
	\left.
	{\partial \epsilon_1(\lambda)\over\partial\lambda}
	\right|_{\lambda=\Lambda_i}\ .
\label{veloF}
\end{equation}

\setlength{\baselineskip}{13pt}

\newpage

\section*{Figures}


\end{document}